\documentclass[pra,twocolumn,longbibliography,nofootinbib]{revtex4-2}

\usepackage[utf8]{inputenc}
\usepackage{bm}
\usepackage{mathtools}
\usepackage{bbold}
\usepackage{hyperref}
\usepackage[normalem]{ulem}
\usepackage{color}

\begin{document}

\title{Robust spin squeezing from the tower of states of $U(1)$-symmetric spin Hamiltonians}
\author{Tommaso Comparin, Fabio Mezzacapo and Tommaso Roscilde}
\affiliation{Univ Lyon, Ens de Lyon, Univ Claude Bernard, CNRS, Laboratoire de Physique, F-69342 Lyon, France}


\begin{abstract}
Spin squeezing -- a central resource for quantum metrology -- can be generated via the non-linear, entangling evolution of an initially factorized spin state. Here we show that robust (i.e. persistent) squeezing dynamics is generated by a very large class of $S=1/2$ spin Hamiltonians with axial symmetry, in relationship with the existence of a peculiar structure of the low-lying Hamiltonian eigenstates -- the so-called Anderson's tower of states. Such states are fundamentally related to the appearance of spontaneous symmetry breaking in quantum systems; and, for models with sufficiently high connectivity, they are parametrically close to the eigenstates of a planar rotor (Dicke states), in that they feature an anomalously large value of the total angular momentum. Our central insight is that, starting from a coherent spin state, a generic $U(1)$-symmetric Hamiltonian featuring  the Anderson's tower of states  generates the same squeezing evolution at short times as the one governed by the paradigmatic one-axis-twisting (or planar-rotor) model of squeezing dynamics. The full squeezing evolution of the planar-rotor model is seemingly reproduced for interactions decaying with distance $r$ as $r^{-\alpha}$ when $\alpha < 5d/3$ in $d$ dimensions. Our results connect quantum simulation with quantum metrology by unveiling the squeezing power of a large variety of Hamiltonian dynamics that are currently implemented by different quantum simulation platforms.    
\end{abstract}
\maketitle


\section{Introduction}
The controlled generation and manipulation of massively entangled quantum states is one of the central tasks of modern quantum technology platforms \cite{Preskill2012Arxiv}. In the context of $S=1/2$ (or qubit) ensembles, a fundamental class of entangled quantum states is represented by spin-squeezed states \cite{Kitagawa1993PRA, Ma2011PR, Pezze2018RMP}, namely states which feature a net polarization of the collective spin, along with suppressed fluctuations of a spin component transverse to the polarization axis. Introducing the collective spin of $N$ qubits, $\hat{\bm J} = \sum_{i=1}^N \hat{\bm S}_i$ -- where $\hat{\bm S}_i$ is a $S=1/2$ spin operator -- and assuming that $\langle \hat J^{y,z} \rangle = 0$ while $\langle \hat J^x \rangle \neq 0$, a state is spin-squeezed if \cite{Wineland1994PRA}
\begin{equation}
\xi_R^2 =
\frac{N \min_{\perp}[{\rm Var}(\hat J^{\perp})]}{\langle \hat J^x \rangle^2}
< 1,
\end{equation}
where $\min_{\perp}$ indicates the minimization over the spin components transverse to the polarization axis $x$. A state exhibiting squeezing is entangled \cite{Sorensen2001}; specifically, if $\xi_R^2 < 1/k$ with integer $k$, the state cannot be represented as a separable state among clusters of less than $k+1$ spins \cite{Pezze2009PRL, Hyllus2012PRA, Toth2012PRA}; moreover the achievement of strong spin squeezing allows a quantum state to violate many-body Bell inequalities, as recently shown experimentally \cite{Schmied2016, Engelsen2017PRL}. Most importantly, spin squeezing offers a fundamental metrological advantage over separable states when used as the input for Ramsey interferometry \cite{Wineland1994PRA} allowing one to beat the standard quantum limit, as already demonstrated in seminal experiments (see e.g. Refs.~\cite{LouchetChauvet2010NJP,Leroux2010PRL,Muessel2014PRL,PedrozoPenafiel2020}). Identifying different many-body mechanisms leading to spin squeezing is therefore of paramount importance: the entanglement of the resulting states can be certified, and their metrological potential exploited by accessing their most basic physical property, namely the collective spin. 

The paradigmatic spin-squeezing dynamics is the one generated by the so-called one-axis-twisting (OAT) Hamiltonian \cite{Kitagawa1993PRA}
\begin{equation}
\hat{\cal H}_{\rm OAT} = \frac{({\hat J}^z)^2}{2I} ~,
\label{e.OAT}
\end{equation}
namely, the Hamiltonian of a planar rotor with moment of inertia $I$.
For the energy to be extensive, one must assume that $I \sim N$. Under this assumption, starting from the initial state $|{\rm CSS}_x\rangle = \otimes_{i=1}^{N}|\rightarrow_x\rangle_i$ (the coherent spin state - CSS - polarized along the $x$ axis) the strongest squeezing is achieved at a time $t_{\rm min} \sim N^{-2/3} I \sim N^{1/3}$; and it scales as $\xi^2_{R,\min} \sim N^{-2/3}$ \cite{Kitagawa1993PRA}.  This Hamiltonian (as well as related ones) and the corresponding squeezing dynamics have been realized in seminal experiments exploiting interactions in Bose-Einstein condensates \cite{Esteve2008, Riedel2010, Hosten2016} and in trapped ions \cite{Bohnet2016}, to cite a few relevant examples.

Interestingly, the OAT Hamiltonian of Eq.~\eqref{e.OAT} plays also a special role in condensed matter, to explain the mechanism of spontaneous breaking of a continuous symmetry in quantum spin models \cite{Beekman2019SPPLN, Lauchli2016, Tasaki2018JSP}.
Hereafter, we consider models with an axial rotational ($U(1)$) symmetry, and we shall choose the $z$ axis as the symmetry axis.
Such models have a ground state which is also $U(1)$ symmetric for any finite size. As suggested by Anderson in pioneering works \cite{Anderson1952PR,Anderson1997}, a quantum spin model in the thermodynamic limit can break a continuous symmetry (such as  $U(1)$)  by developing a finite order parameter due to the existence of a set of low-energy Hamiltonian eigenstates -- the so-called Anderson's tower of states (ToS). These states are approximately eigenstates of a planar rotor Hamiltonian of the kind of Eq.~\eqref{e.OAT}; and their energy decreases as ${\cal O}(1/N)$ in the thermodynamic limit (due to the scaling of the moment of inertia $I$), making them nearly degenerate with the $U(1)$ symmetric ground state. Spontaneous symmetry breaking (SSB) is therefore the result of the collapse of the ToS. 

The central insight of the present work is that generic quantum spin Hamiltonians can produce a persistent spin-squeezing dynamics thanks to the emergent ToS structure of their low-energy eigenstates -- in other words, several models with $U(1)$ symmetry generate the same dynamics as that of the OAT Hamiltonian Eq.~\eqref{e.OAT} at sufficiently short times, irrespective of the nature (long-range vs. short-range) of their interactions.
This is strictly true when the dynamics is initialized in a CSS polarized in the symmetry plane. Such a state, maximizing the total spin $\langle \hat {\bm J}^2 \rangle = J_{\rm max}(J_{\rm max}+1)$ with $J_{\rm max} = N/2$ (for an ensemble of qubits), has a very strong overlap with the ToS, because the latter states have in turn an anomalously large average value of  $\hat {\bm J}^2$ among all the states in the same energy range -- namely they behave as ``quantum many-body scars" \cite{Turner2018NP, Serbyn2021NP}, whose presence alters profoundly the dynamics of the system with respect to a generic initial state with the same initial energy. 
We underpin this insight with a combination of exact diagonalization (ED) and time-dependent variational calculations of the dynamics of various $S=1/2$ $XX$ models, fully corroborating the universal picture of spin-squeezing dynamics. 
Our results imply that a large variety of current quantum simulation setups implementing $U(1)$ symmetric models of quantum magnetism \cite{Brydges2019, Browaeys2020NP, Jepsen2020, Chiaro2022PRR, Guo2021} can be viewed as generators of spin-squeezed states, of potential immediate interest for quantum metrology tasks.

{The structure of the paper is as follows. In Sec.~\ref{section:model_method} we briefly present the $\alpha-XX$ model and our numerical approach to simulate quantum dynamics; in Sec.~\ref{section:tos} we identify the Tower of States for the model under study; Sec.~\ref{section:squeezing} discusses our variational results for spin squeezing and Sec.~\ref{section:squeezing_and_tos} describes the connection between the robustness of squeezing dynamics and the Tower of States.}

\section{Models and method}
\label{section:model_method}

We specialize our attention to the case of $XX$ models with power-law decaying interactions ($\alpha-XX$ models), whose Hamiltonian reads
\begin{equation}
\hat{\cal H} = - \sum_{i<j} {\cal J}^{(\alpha)}_{ij}  \left ( \hat S_i^x \hat S_j^x + \hat S_i^y \hat S_j^y \right )  ~,
\end{equation} 
where ${\cal J}^{(\alpha)}_{ij} =  {\cal J} \, |\bm r_i - \bm r_j|^{-\alpha}$ with $\alpha \geq 0$. The limit $\alpha = 0$ corresponds to infinite-range interactions (reproducing the OAT Hamiltonian), while the opposite limit $\alpha \to \infty$ corresponds to nearest-neighbor interactions~\footnote{The $\alpha-XX$ Hamiltonian features extensive eigenvalues only for $\alpha>d$, $d$ being the number of dimensions; later, when appropriate, we will adopt a Kac normalization of the coupling constant in order to reinstate the extensive nature of the energy.}.
The above Hamiltonian occupies a prominent role in quantum simulation,  as it is currently implemented on various lattice geometries by a rich variety of very different quantum simulation platforms, including trapped ions ($0 < \alpha < 3$) \cite{Brydges2019}; Rydberg atoms ($\alpha = 3$) \cite{Browaeys2020NP}; spinor gases ($\alpha = \infty$) \cite{Jepsen2020}; and superconducting circuits ($\alpha = \infty$) \cite{Chiaro2022PRR, Guo2021}.  The generation and metrological exploitation of spin squeezed states in these platforms has been already discussed in the recent past (see e.g. Refs.~\cite{Gil2014PRL, Bohnet2016, Qu2019PRA, VanDamme2021PRA, Groszkowski2022PRX}), although not in connection with the operation mode implementing the $\alpha-XX$ model. 

Throughout the paper, we will consider the quench dynamics generated by $\alpha-XX$ Hamiltonians starting from the  $|{\rm CSS}_x\rangle$ state, $|\Psi(t)\rangle = \exp(-i\hat{\cal H} t) |{\rm CSS}_x\rangle$. Under this assumption, the sign of the ${\cal J}$ coupling is irrelevant as long as one follows the expectation value of operators which are real matrices (e.g. on the computational basis of the eigenstates of the $\hat{S}_i^z$ operators) \cite{Frerot2018PRL}. Hereafter we will assume ${\cal J} >0$ for definiteness.  We shall study this model on lattices with $N=L^d$ sites (in $d=1$ and 2) with periodic boundary conditions. 

Studying the real-time dynamics generated by the Hamiltonian with $\alpha > 0$ is generically a challenging problem. 
Recent results based on a semi-classical approach \cite{Perlin2020PRL} already indicate the robustness of squeezing when moving away from the $\alpha=0$ limit. Here we adopt a different strategy, going beyond any semi-classical framework.
Specifically, we use exact diagonalization for small systems \cite{Weinberg2017SP, Weinberg2019SP}, and a time-dependent variational Monte Carlo (tVMC) approach to tackle large system sizes. The latter are both relevant to current experimental realizations and instrumental to our scaling analysis of the spin-squeezing dynamics  discussed below.
Our tVMC calculations are based on the pair-product Ansatz (or two-spin long-range entangled-plaquette state, 2LR-EPS \cite{Thibaut2019PRB})
\begin{equation}
|\Psi(t)\rangle =:
\sum_{\bm \sigma} \prod_{i<j} \psi_{ij}(\sigma_i,\sigma_j;t) |\bm \sigma\rangle,
\label{e.ansatz}
\end{equation}
where $|\bm \sigma \rangle = |\{ \sigma_i \}\rangle$ is the joint eigenstate of all $\hat{S}_i^z$ operators. The evolution of the variational parameters $\psi_{ij}(\sigma_i,\sigma_j;t)$ is dictated by the time-dependent variational principle, and it requires Monte Carlo sampling of the probability distribution $|\langle {\bm \sigma} | \Psi(t) \rangle|^2$ \cite{Carleo2012SR, Becca2017}.
{As discussed in the following,} the chosen Ansatz has the advantage of reproducing the \emph{exact} dynamics in the $\alpha=0$ limit; and, as we shall see, it remains very accurate for $\alpha > 0$. 

{\subsection{Pair-product Ansatz is exact for the one-axis twisting Hamiltonian}

The exact evolution of the CSS under the OAT Hamiltonian is restricted to the $J = J_{\rm max}$ sector, and it reads
\begin{equation}
\begin{aligned}
&e^{-i \hat{\cal{H}}_{\rm OAT} t} |{\rm CSS}_x \rangle
=\\
&\sum_{J^z=-J_{\rm max}}^{J_{\rm max}}
\langle J_{\rm max}, J^z | {\rm CSS}_x \rangle
\exp\left[-it \frac{(J^z)^2}{2 I} \right] |J_\mathrm{max}, J^z \rangle,
\end{aligned}
\end{equation}
where we set $\hbar = 1$ and where
\begin{equation}
\langle J_{\rm max}, J^z | {\rm CSS}_x \rangle =
2^{-J_{\rm max}}
\sqrt{\begin{pmatrix} 2 J_{\rm max} \\ J_{\rm max}-J^z \end{pmatrix} } ~.
\end{equation}
Provided that a certain variational Ansatz correctly represents the initial state, then it can also exactly reproduce the time-evolved state if the coefficients $\langle {\bm \sigma} | \Psi(t) \rangle$ can take values proportional to $\exp[-i t (J^z)^2 / (2 I)]$, with $J^z = \sum_{i=1}^N \sigma_i$ and where $\sigma_i = \pm 1/2$ is the eigenvalue of $\hat{S}_i^z$.
The pair-product Ansatz in Eq.~\eqref{e.ansatz} has this property, as we show by construction.
If we set its coefficients to be
\begin{equation}
\psi_{jk}(\sigma_j, \sigma_k; t) =
\exp\left[w_{jk} \sigma_j \sigma_k \right] ~,
\qquad
w_{jk} = -\frac{i t}{I} ~,
\label{se.ansatz_for_OAT}
\end{equation}
then the value of $|\Psi(t)\rangle$ on a given basis state $|{\bm \sigma}\rangle$ reads
\begin{equation}
\begin{aligned}
\langle {\bm \sigma} | \Psi(t) \rangle &=
\prod_{j<k} \psi_{jk}(\sigma_j, \sigma_k; t) =
\exp
\left[ -\frac{i t}{I} \sum_{j<k}  \sigma_j \sigma_k \right]
=\\
&=
\exp\left[-i t \, \frac{(J^z)^2}{2I} \right]
\exp\left[ i t \, \frac{N}{8I} \right].
\end{aligned}
\end{equation}
This corresponds to the required form of $\langle {\bm \sigma} | \Psi(t) \rangle$, up to an irrelevant global phase factor.
Therefore the pair-product Ansatz can describe the exact time evolution of $|{\rm CSS}_x\rangle$. Note that the initial state is trivially represented with this Ansatz by setting all coefficients $\psi_{jk}(\sigma_j, \sigma_k ; t = 0)$ equal to each other.

The Hamiltonian of the $\alpha-XX$ model with $\alpha=0$,
\begin{equation}
\hat{\cal{H}} =
\frac{\mathcal{J}}{2} \left[(\hat{J}^z)^2 - \hat{\bm J}^2\right] + \mathrm{const}~,
\end{equation}
corresponds to the OAT Hamiltonian with moment of inertia $I = 1 / {\cal J}$, up to a constant shift and to an additional $\hat{\bm J}^2$ term (which only adds an irrelevant global phase factor to the time-evolved state).
Thus the pair-product Ansatz is also exact for the $\alpha-XX$ model with $\alpha = 0$.
We explicitly verify this fact by comparing the tVMC dynamics of spin squeezing to the exact expression for the OAT squeezing dynamics \cite{Kitagawa1993PRA}. As shown in Fig. \ref{f.jastrow_OAT}, the tVMC calculation for the $\alpha-XX$ model with $\alpha=0$ is exact.
Note that the dynamics of variational parameters is obtained through the tVMC scheme \cite{Becca2017} {based on the time-dependent variational principle}, that is, without postulating the expected expression in Eq. \eqref{se.ansatz_for_OAT}.

\begin{figure}[hbt]
\centering
\includegraphics[width=0.8\linewidth]{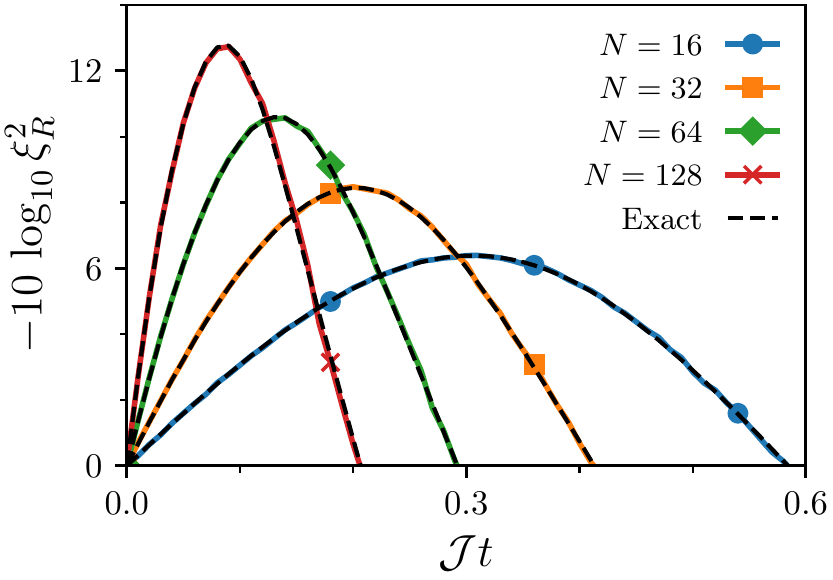}
\caption{Spin squeezing generated by the $\alpha-XX$ Hamiltonian, with $d=1$ and $\alpha=0$. For each system size $N$, the result of tVMC with the pair-product Ansatz (solid lines) perfectly matches with the exact solution for the OAT Hamiltonian \cite{Kitagawa1993PRA}.}
\label{f.jastrow_OAT}
\end{figure}

}

\section{Tower of states as quantum many-body scars}
\label{section:tos}

\begin{figure*}[ht!]
\includegraphics[width=0.95\textwidth]{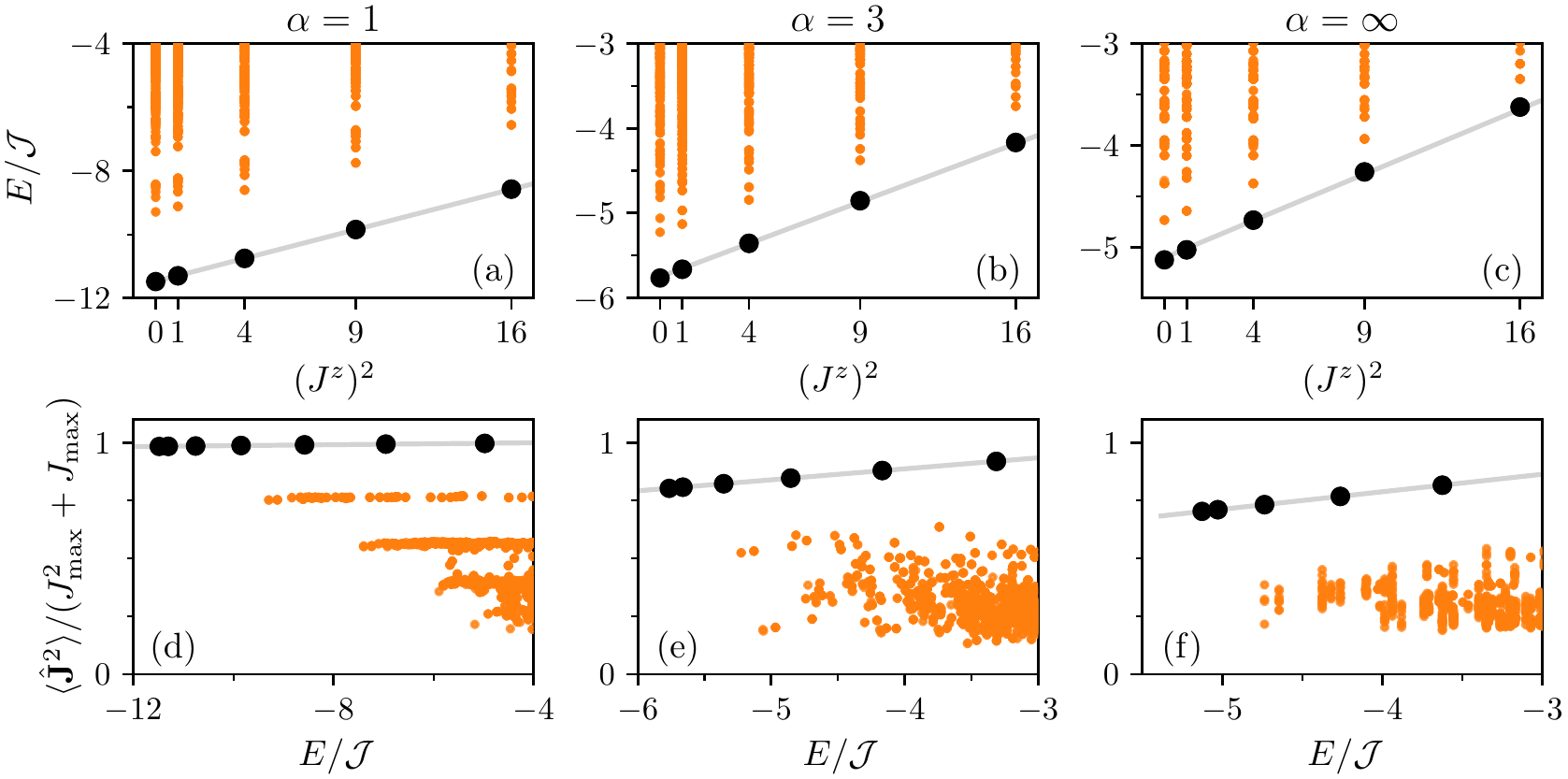}
\caption{Low-energy spectrum of the $d=1$ $\alpha-XX$ model for (a)-(d) $\alpha = 1$; (b)-(e) $\alpha = 3$; (c)-(f) $\alpha = \infty$. The spectrum is shown as a function of the two quantum numbers $E = \langle \hat{\cal H} \rangle$  and $(J^z)^2$ in panels (a-c), and as a function of $E$ and the average total spin $\langle \hat {\bm J}^2 \rangle$ in panels (d-f). Black dots correspond to states of the ToS, while orange (gray) dots are outside the ToS; straight lines are linear fits of the ToS states.}
\label{f.ToS}
\end{figure*}

The existence of an Anderson's ToS in the low-energy spectrum of the $\alpha-XX$ model can be directly verified using ED on small system sizes \cite{Lauchli2016}. Fig.~\ref{f.ToS} shows the low-energy spectrum of a $N=16$ chain with $\alpha = 1, 3$ and $\infty$; the ToS is clearly visible when plotting the energies as a function of the quantum number $J^z$, since the ground states in each sector have an energy which grows almost exactly as $(J^z)^2$, as expected for the eigenstates of $\hat{\cal H}_{\rm OAT}$.  Remarkably, a ToS is also observed for $\alpha \geq 3$, although in this case it does not collapse onto the ground state faster than the spin-wave excitations do \cite{Frerot2017PRB}, so that SSB is not realized in the ground state \cite{Maghrebi2017PRL}.

The states of the ToS, being ground states of different $J^z$ sectors, can be naturally expected to exhibit non-typical features among all the states in the corresponding energy range. In particular, when mapping the $\alpha-XX$ model onto hardcore bosons \cite{Matsubara1956PTP}, $J^z$ parametrizes the particle number, such that the ToS is readily understood as the set of the (quasi-)condensate ground states for different particle numbers~\footnote{The states of the ToS, being ground states of hardcore bosons with long-range hopping, are expected to exhibit area-law scaling of entanglement entropy, at least for $\alpha > d$, while logarithmic scaling of the entropy with subsystem size can be exactly proven for $\alpha=0$. This is in stark contrast with typical states at finite energy density.}. In particular, the ToS contains the states featuring the most slowly decaying correlations $C^{x(y)}_{ij} = \langle \hat S_i^{x(y)} \hat S_j^{x(y)} \rangle$ (corresponding to the one-body density matrix for bosons) among all the states at the same $J^z$: therefore they are the states maximizing the total angular momentum $\langle \hat {\bm J}^2 \rangle = (J^z)^2 + \sum_{ij}  C^{x(y)}_{ij}$. Fig.~\ref{f.ToS} shows that they are also the states with the largest $\langle \hat{\bm J}^2 \rangle$ throughout their energy range. As such, they represent a paradigmatic example of  quantum many-body scars \cite{Turner2018NP}.
Due to their anomalously large value of $\langle \hat{\bm J}^2 \rangle$,  the states of the ToS  are naturally expected to play a significant role in the dynamics of the system when starting from the $|{\rm CSS}_x\rangle$ state, given that for this state $\langle \hat{\bm J}^2 \rangle$ is maximal.  Indeed, as shown in Appendix~\ref{app:overlap}, the CSS has maximal overlap with the ToS. 

\section{Squeezing dynamics and its scaling}
\label{section:squeezing}

\begin{center}
\begin{figure*}[htb]
\includegraphics[width=0.75 \textwidth]{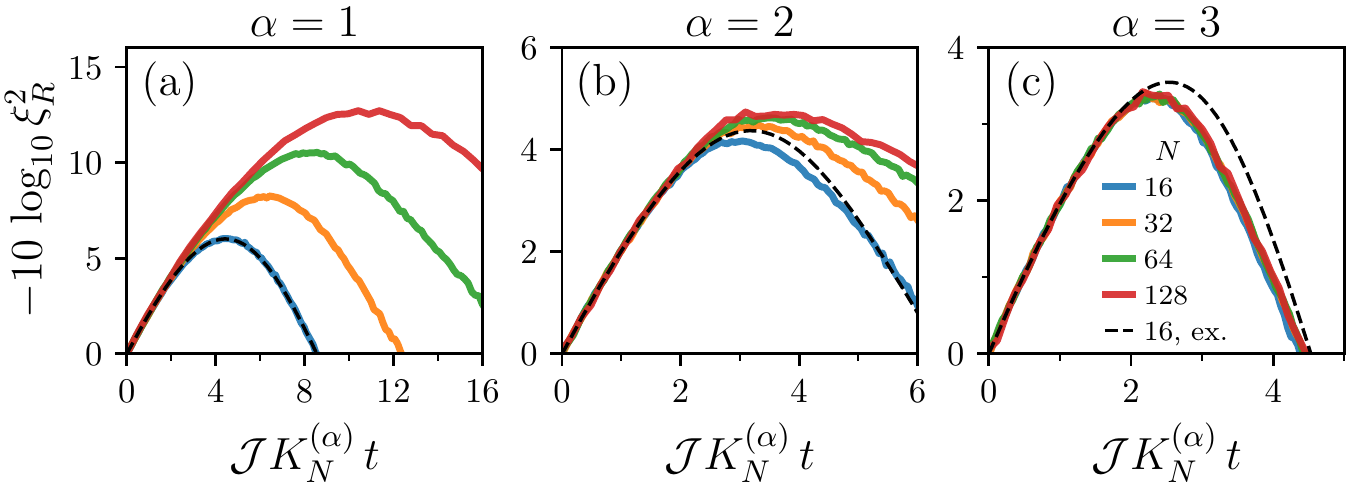}
\caption{Dynamics of the squeezing parameter for the 1$d$ $\alpha-XX$ Hamiltonian with $\alpha = 1$ (a), $\alpha = 3$ (b) and $\alpha = \infty$ (c), for system size from $N=16$ (bottom line) to $N=128$ (top line).
Time is rescaled by the Kac prefactor $K_N^{(\alpha)}$ -- see text. The dashed line is the exact result for $N=16$, while the other data are obtained via tVMC (solid lines).
}
\label{f.squeezing}
\end{figure*}
\end{center}

The existence of a set of quantum-scar states with an energy dependence on $J^z$ akin to that of the OAT Hamiltonian in Eq.~\eqref{e.OAT}, and which are strongly overlapping with the CSS, suggests that, at sufficiently short times, the Hamiltonian evolution starting from the CSS will strongly resemble that of the OAT model -- namely it will feature squeezing. Probing the persistence of this squeezing dynamics and its quantitative relationship with the Anderson ToS will be the goal of the rest of this work.
Fig.~\ref{f.squeezing} clearly exhibits the existence of squeezing dynamics for different $\alpha$ values in $d=1$ and for variable system sizes up to $N=128$ spins. Our tVMC results are indistinguishable from the exact ones (which we obtain for $N=16$) up to $\alpha \approx 1$, remaining accurate for all values of $\alpha$ -- see also additional data in Appendix~\ref{app:squeezing_data}.

\begin{figure}[ht!]
\centering
\includegraphics[width=\linewidth]{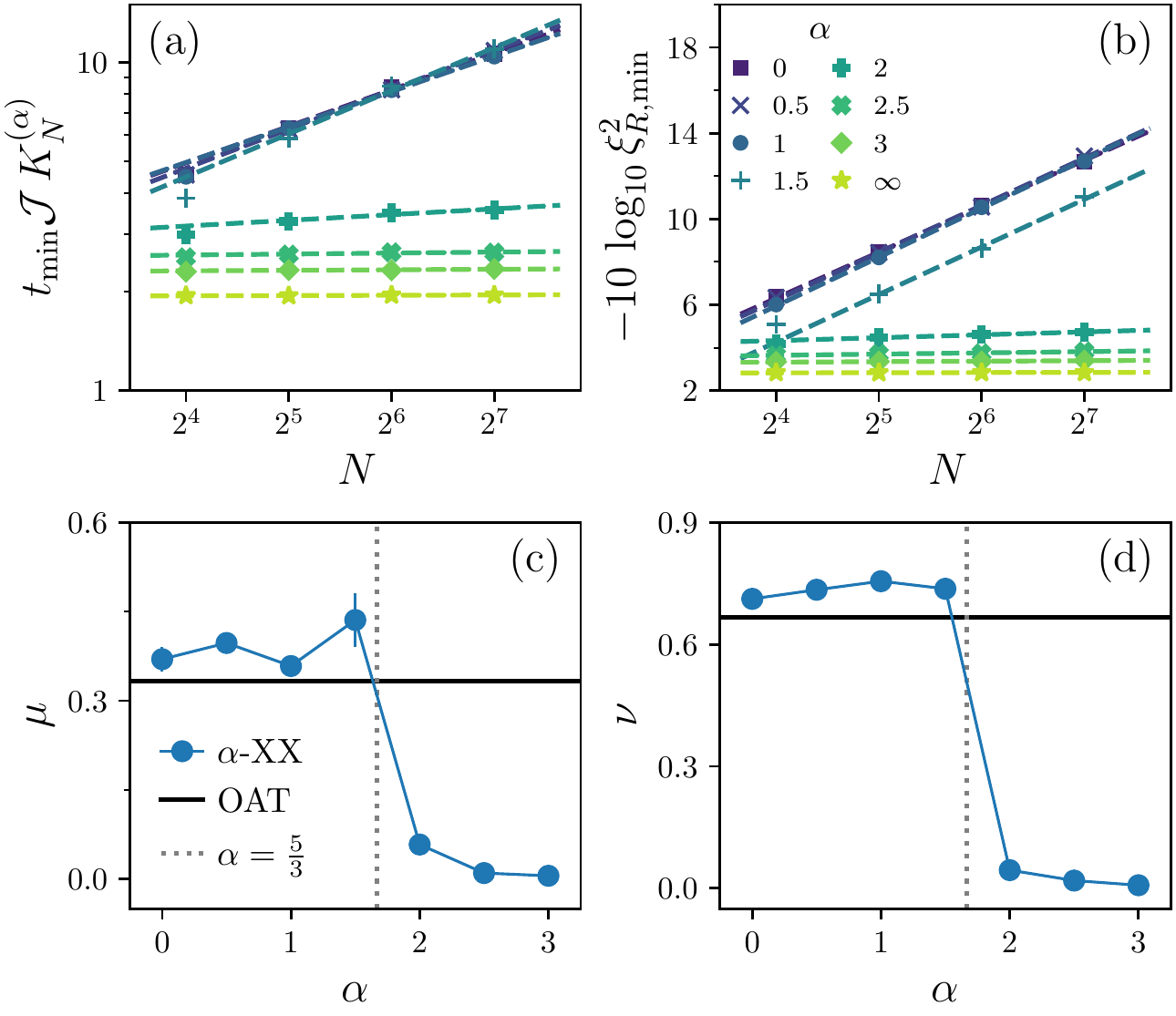}
\caption{Finite-size scaling of the optimal time $t_\mathrm{min}$ (a) and of the minimal squeezing $\xi^2_{R,\mathrm{min}}$ (b), for several values of $\alpha$. Dots are tVMC results; straight lines are power-law fits with $t_\mathrm{min}\mathcal{J} K_N^{(\alpha)} \sim N^{\mu}$ and $\xi^2_{R,\mathrm{min}} \sim N^{-\nu}$. The fit range is $32 \leq N \leq 128$.
(c) Exponent $\mu$ for the scaling of the optimal time. (d) Exponent $\nu$ for the scaling of the minimum squeezing parameter. In panels (c-d), the horizontal black line shows the OAT exponents, and the vertical dotted line marks $\alpha=5d/3$ (see text).
}
\label{f.squeezing_scaling}
\end{figure}

{ For each $\alpha$, we identify the minimal value of the squeezing parameter $\xi^2_{R,\min}$ and the corresponding optimal time $t_{\rm min}$;
Figs. \ref{f.squeezing_scaling}(a) and \ref{f.squeezing_scaling}(b) show the scaling of these two quantities with the system size.}
The results of the scaling analysis in the $1d$ $\alpha-XX$ model are summarized in Fig.~\ref{f.squeezing_scaling}(c-d). Following the example of the OAT model, we postulate \cite{FossFeig2016Arxiv, Perlin2020PRL} the power-law scalings 
\begin{equation}
\xi^2_{R,\min} \sim N^{-\nu},  ~~~~~~~ t_{\rm min} K^{(\alpha)}_N \sim N^{\mu}~. 
\end{equation}
Here we have introduced the Kac normalization 
\begin{equation}
K^{(\alpha)}_N = N^{-1} \sum_i \sum_{j \neq i} |\bm r_i - \bm r_j|^{-\alpha}
\end{equation}
of the power-law couplings in order to properly scale time, by considering evolutions with Hamiltonians with extensive energies~\footnote{In particular, in the limit $N\to\infty$, we have that  $K^{(\alpha)}_N \sim N^0$ for $\alpha > 1$, $K^{(\alpha)}_N \sim \ln N$ for $\alpha = 1$ and $K^{(\alpha)}_N \sim N^{1-\alpha}$ for $\alpha < 1$.} -- this is appropriate for all the quantum-simulation platforms cited above. The system sizes we have considered (up to $N=128$) do not necessarily capture the asymptotic scaling limit, but they are comparable with the typical sizes achieved by state-of-the-art quantum simulators for the $\alpha-XX$ model. 
In Fig.~\ref{f.squeezing}(d-e) we observe that the scaling properties of the OAT limit $\alpha = 0$ (namely $\mu = 1/3$ and $\nu = 2/3$) are essentially maintained throughout the range $0 < \alpha \lesssim 1.5$. This observation reveals (far beyond what is accessible to ED) the dominant role that the ToS has on the dynamics of the system in this range of $\alpha$, exceeding the regime of long-range interactions $\alpha \leq 1$. On the other hand for $\alpha \gtrsim 2$ both $\mu$ and $\nu$ are found to sharply drop to zero; this observation signals that the dynamics leaks significantly out of ToS manifold, and that further Hamiltonian eigenstates entering in the dynamics have the effect of curbing the growth of squeezing and suppressing its scaling behavior. {In the following section,  using spin-wave (SW) theory \cite{Frerot2017PRB,Frerot2018PRL}, we can estimate that $\alpha < 5d/3$ (for $d\leq 3$) is a necessary condition for the squeezing dynamics of the OAT to be reproduced at $\alpha > 0$. This would translate to $\alpha < 1.6666...$ in $d=1$, in apparent agreement with the observation of Fig.~\ref{f.squeezing_scaling}(d).}

{ \subsection{Necessary condition on the persistence of $\alpha=0$ scaling of squeezing from spin-wave theory }

In a (Kac-normalized) OAT Hamiltonian, with all-to-all interactions, the optimal squeezing time scales as 
\begin{equation}
t_{\rm min} \sim N^{1/3} = L^{d/3}~.
\end{equation}
Taking the OAT Hamiltonian literally, its dynamics is characterized by the total absence of retardation effects: any signal propagates instantaneously from end to end of the system. As a consequence, the dynamics of the spin system is completely captured by that of the collective spin variables, and there is no relative dynamics between spins.  
Even in the picture of the OAT model, there is still a finite time required for the establishment of maximal squeezing, scaling with system size. This scaling also contains the Kac normalization, namely the interactions have no retardation, but their strength is decreasing as $N^{-1}$ to keep the energy extensive. This has the fundamental effect of making the optimal time increase with system size. 
 
When considering instead $\alpha > 0$, the relevant excitations involved in the dynamics starting from the CSS$_{x}$ state are not only the states from Anderson's ToS, but also spin-wave (SW) excitations. These excitations may have in general a finite group velocity, or a group velocity diverging with system size sufficiently slowly, so that they could lead to retardation effects in the dynamics of the collective spin, thereby altering substantially the picture of the OAT model.  
 
SW theory for the $\alpha-XX$ model \cite{Frerot2017PRB,Frerot2018PRL} predicts that SW excitations have a dispersion relation $\omega \sim k^z$  at small $k$, with dynamical exponents $z$ taking values
\begin{equation}
z = \begin{cases} 1  ~~~~~~~~~~~ \alpha \geq d+2 \\ \frac{\alpha - d}{2}  ~~~~~~~ d \leq \alpha \leq d+2 \\ 0 ~~~~~~~~~~~ \alpha \leq d \end{cases}~.
\end{equation}
The associated group velocity is therefore $v_g \sim k^{z-1}$. On a finite system of linear size $L$ the maximum group velocity is therefore scaling as $v_{g, {\rm max}} \sim k_{\rm min}^{z-1} \sim L^{1-z}$ where $k_{\rm min} = 2\pi/L$. Associated with this maximal group velocity there is an intrinsic minimal retardation time 
\begin{equation}
t_{\rm SW} = \frac{L}{v_{g,\rm max}} \sim L^z
\end{equation}
which is the time needed for the fastest spin-wave excitations to traverse the entire system of linear size $L$. 

When $\alpha \leq d$, the vanishing of the $z$ exponent implies that the retardation time does not scale with system size, so that the spatial decay of interactions is not expected to affect the collective spin dynamics at the time scale of $t_{\rm min}$ (which instead grows with $L$). On the other hand, when $\alpha > d$ one has that $z>0$, so that retardation effects can play a role. Given that $t_{\rm SW}$ is the minimal retardation time in the collective spin dynamics, one can argue that a \emph{necessary} condition for retardation effects not to affect the collective spin dynamics up to the optimal squeezing time $t_{\rm min}$ is that 
\begin{equation}
t_{\rm SW} < t_{\rm min}~.    
\end{equation}
For this condition to remain valid for large system sizes, one should then require that $L^z < L^{d/3}$, namely
\begin{equation}
z <  \frac{d}{3} ~~ \Longrightarrow ~~ \alpha < \frac{5}{3} d  ~.    
\end{equation}
The above necessary condition is valid if $\frac{5}{3} d \leq d+2$, namely if $d \leq 3$, covering all situations of interest. 
For $d=1$, the necessary condition reads $\alpha<5/3=1.666..$, and our numerical results suggest that this necessary condition may also be sufficient -- see Fig.~\ref{f.squeezing_scaling}(c-d).
{If the same condition is sufficient as well for $d>1$, this would imply that the OAT scaling of spin squeezing dynamics is expected e.g. in dipolar systems ($\alpha = 3$) in $d=2$, a situation which is highly relevant for experiments on arrays of Rydberg atoms \cite{Browaeys2020NP} or for dipolar molecules in optical lattices \cite{Yan2013}, to cite two important examples.}

{We also remark that, in the case of the $\alpha-XX$ model in $d=1$ discussed here, the persistence of the OAT squeezing dynamics for $\alpha < 5/3$ is accompanied by the persistence of a large value of the total spin length, $\langle {\bm J}^2 \rangle$ for the largest system sizes we considered, as shown in  Appendix~\ref{app:Jsquare}; while for larger $\alpha$ values $\langle {\bm J}^2 \rangle$ decays to values which decrease rather fast with system size. This aspect must also be related to the $\alpha$-dependence of the low-temperature thermodynamics of the system, which in principle dictates the long-time behavior of the non-equilibrium evolution within a picture of thermalizing dynamics, expected for the $\alpha-XX$ chain which is non-integrable for all $0 < \alpha < \infty$. Establishing a quantitative relationship is nonetheless beyond the scope of the current work, and it will be subject of a future work.}  

\begin{center}
\begin{figure}[ht!]
\includegraphics[width=\columnwidth]{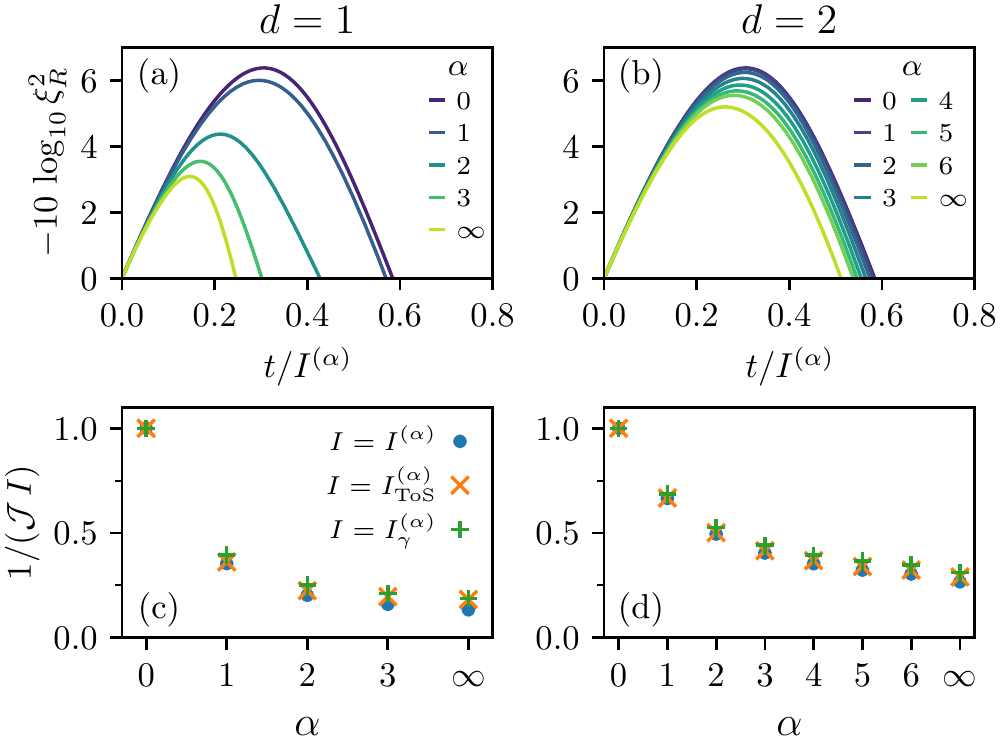}
\caption{\emph{Universal short-time squeezing dynamics.} Squeezing dynamics for the $\alpha-XX$ model in (a) $d=1$ ($N=16$) and (b) $d=2$ ($N=4\times 4$), for various values of $\alpha$ from $\alpha=0$ (darkest line) to $\alpha=\infty$ (lightest line). Collapse of the curves at short times is obtained by rescaling time with the $I^{(\alpha)}$ parameter;  (c-d) Comparison between the three estimates of the moment of inertia: from the Hamiltonian projection onto Dicke states ($I^{(\alpha)}$); from the Fourier transform of the coupling constant ($I_{\gamma}^{(\alpha)}$); and from the ToS spectrum ($I_{\rm ToS}^{(\alpha)}$). System Hamiltonian and sizes as in panels (a-b).}
\label{f.universal}
\end{figure}
\end{center} 
}

\section{Relationship of the squeezing dynamics to the ToS spectrum}
\label{section:squeezing_and_tos}

The connection between the OAT dynamics and the short-time dynamics of generic $\alpha-XX$ models can be made fully quantitative by identifying the moment of inertia $I^{(\alpha)}$ of the effective OAT Hamiltonian emerging from the $\alpha-XX$ model. The general way to do this is to assume that the short-time dynamics of the $\alpha-XX$ model prepared in the CSS remains confined in the sector with maximum total angular momentum $J_{\rm max}$ -- this assumption will be justified below. If this is true, then the effective Hamiltonian governing the dynamics is given by the restriction of the $\alpha-XX$ model onto the subspace of Dicke states $|J_{\rm max},J^z\rangle$ which are eigenstates of $\hat{\bm J}^2$ with $J = N/2$, as well as of $\hat J^z$; given that $\hat {\cal H}$ commutes with $\hat J^z$ and it is invariant under inversion of the spins along $z$, the resulting projected Hamiltonian is necessarily an even function of $J^z$
\begin{eqnarray}
 \langle J_{\rm max}, J^z | \hat {\cal H} | J_{\rm max}, (J^z)' \rangle  = ~~~~~~~~~~~~~~~~~~~~~&&\nonumber \\
  \delta_{J^z, (J^z)'} [ {\rm const.} + (J^z)^2/(2I^{(\alpha)}) + {\cal O}(J^z)^4]~. &&
 \end{eqnarray}
  The eigenvalues of the Hamiltonian projected onto the Dicke states turn out to be almost perfect quadratic functions of $J^z$ (Appendix \ref{app:diagonal}), which allow us to systematically extract the corresponding moment of inertia $I^{(\alpha)}$. The picture of the effective dynamics projected onto the Dicke-state manifold would then predict that all $\alpha-XX$ models squeeze the fluctuations of the CSS in the same way as the OAT Hamiltonian does  --  so that universal dynamics should be manifested when properly rescaling time by $t_{\alpha} = I^{(\alpha)}$. This is indeed observed in Fig.~\ref{f.universal}(a-b), exhibiting a perfect collapse for the short-time evolution of the squeezing parameter. Yet the time over which the $\alpha-XX$ model reproduces the spin squeezing dynamics of the OAT is strongly dependent on $\alpha$ and on $d$, as a result of the properties of the ToS in the Hamiltonian spectrum.
 
To quantitatively connect the universal spin-squeezing dynamics observed above with the existence of the ToS in the spectrum one can adopt a Fourier decomposition of the $\alpha-XX$ model \cite{Beekman2019SPPLN} as $\hat {\cal H} = \sum_{\bm q} \hat{\cal H}_{\bm q} + {\rm const.}$, where
 \begin{equation}
\hat{\cal H}_{\bm q} =
-\frac{\cal J}{2N}
(\gamma_{\bm q, N}^{(\alpha)}+1)
\left ( \hat J_{\bm q}^x \hat J_{-\bm q}^x + \hat J_{\bm q}^y \hat J_{-\bm q}^y \right)~.
\end{equation}  
We have introduced the Fourier transform of the collective spin
\begin{equation}
\hat J_{\bm q}^{x(y)} = \frac{1}{\sqrt{N}} \sum_i e^{i\bm q \cdot {\bm r}_i} \hat S^{x(y)}_i~,
\end{equation}
while
\begin{equation}
\gamma^{(\alpha)}_{\bm q, N} = \sum_{\bm r \neq 0} e^{i\bm q \cdot {\bm r}} |\bm r|^{-\alpha}
\end{equation}
is the Fourier transform of the couplings calculated on an $N$-site system. The $\hat{\cal H}_{{\bm q}=0}$ Hamiltonian corresponds to a OAT model with $[I_\gamma^{(\alpha)}]^{-1} = {\cal J}(\gamma^{(\alpha)}_{0,N}+1)/N$.
The emergence of a ToS in the spectrum of the $\alpha-XX$ model can be understood as the fact that the $\hat{\cal H}_{\bm q\neq 0}$ terms in the Hamiltonian perturb only weakly the eigenstates of the ${\cal H}_{{\bm q}=0}$ part with maximal $\langle \hat{\bm J}^2 \rangle$; as a consequence there exist special Hamiltonian eigenstates which are parametrically closer to Dicke states with $J=J_{\rm max}$, the smaller $\alpha$ is.
 These states have energies $E \approx (J^z)^2/(2 I^{(\alpha)}_{\rm ToS}) + {\rm const.}$, with a moment of inertia  $I^{(\alpha)}_{\rm ToS}$ that can be extracted by fitting the energy spectrum as in Fig.~\ref{f.ToS}. The $I^{(\alpha)}_{\rm ToS}$ parameter differs from the $I_\gamma^{(\alpha)}$ parameter in that it is renormalized by the $\hat{\cal H}_{\bm q\neq 0}$ perturbations; and it differs from the $I^{(\alpha)}$ parameter, because it includes the mixing of the $J_{\rm max}$ sector of Hilbert space with all the other $J$ sectors. Yet these three definitions of the effective moment of inertia -- a dynamical one ($I^{(\alpha)}$), an Hamiltonian one  ($I_{\gamma}^{(\alpha)}$) and a spectral one ($I^{(\alpha)}_{\rm ToS}$) -- coincide for $\alpha = 0$, and they turn out to be very close to each other for all $\alpha$ values, as shown in Fig.~\ref{f.universal}(c-d) for both $d=1$ (for $\alpha \lesssim 2$) and $d=2$. This result corroborates the picture in which the $\alpha-XX$ Hamiltonians with high connectivity couple weakly the $J_{\rm max}$ sector with all the other sectors of Hilbert space, resulting simultaneously in  1) the existence of quantum scars (the ToS) in the spectrum, predominantly overlapping with the $J_{\rm max}$ sector; and  2) the existence of robust spin-squeezing dynamics at short times, which remains temporarily confined in the $J_{\rm max}$ sector (see Appendix~\ref{app:Jsquare} for the dynamics of $\langle \hat{{\bm J}}^2 \rangle$), reproducing the behavior of the OAT model with an effective moment of inertia related to the energy spectrum of the ToS.

{ This picture is to be contrasted with the one for models lacking a ToS.
As we showed recently, short-time spin-squeezing dynamics starting from a coherent spin state is a very general phenomenon common to a large class of Hamiltonians featuring bilinear spin-spin interactions with parity conservation~\cite{Roscilde2021PRA}. The ability of an Hamiltonian to induce spin squeezing at short times has therefore  \emph{a priori} no specific relationship to the existence of an Anderson ToS in the spectrum.
 {In Appendix~\ref{app:dimerized} we describe in detail an example (the dimerized $XX$ chain) where the ToS is absent. While short-time squeezing dynamics is also observed for that Hamiltonian (in compliance with the theorem of Ref.~\cite{Roscilde2021PRA}), we show that its characteristic time scale cannot be directly related to features of the spectrum, because the states participating in the dynamics (namely overlapping significantly with the CSS) lie in the ``bulk" of the spectrum and do not form a well-defined tower of quantum scars.} }

\section{Conclusions}

Focusing on $U(1)$ symmetric models, in this work we have unveiled a dynamical manifestation of the Anderson's tower of states -- the fundamental spectral mechanism of spontaneous breaking of continuous symmetries in quantum mechanics -- in the form of an effective squeezing dynamics of coherent spin states, fully equivalent at short times to that generated by a planar-rotor Hamiltonian.
In particular we find that the one-dimensional $XX$ model with power-law interactions generates squeezing that scales with system size when the decay exponent takes values $\alpha \lesssim 5/3$.
These results establish an important link between the simulation of non-equilibrium dynamics of $U(1)$ quantum spin Hamiltonians -- a common task of nearly all quantum simulation platforms -- and quantum metrology.

\begin{acknowledgments} We acknowledge useful discussions with Johannes Schachenmayer. Exact results for small systems are obtained via the QuSpin package \cite{Weinberg2017SP,Weinberg2019SP}.
This work is supported by ANR (``EELS" project) and by QuantERA (``MAQS" project). All numerical simulations have been performed on the PSMN cluster of the ENS of Lyon. Data and additional details about the tVMC numerical simulations are made publicly available \cite{Comparin2022PRA_suppdata}.
\end{acknowledgments}


\appendix

\begin{figure*}[hbt!]
\centering
\includegraphics[width=0.9\linewidth]{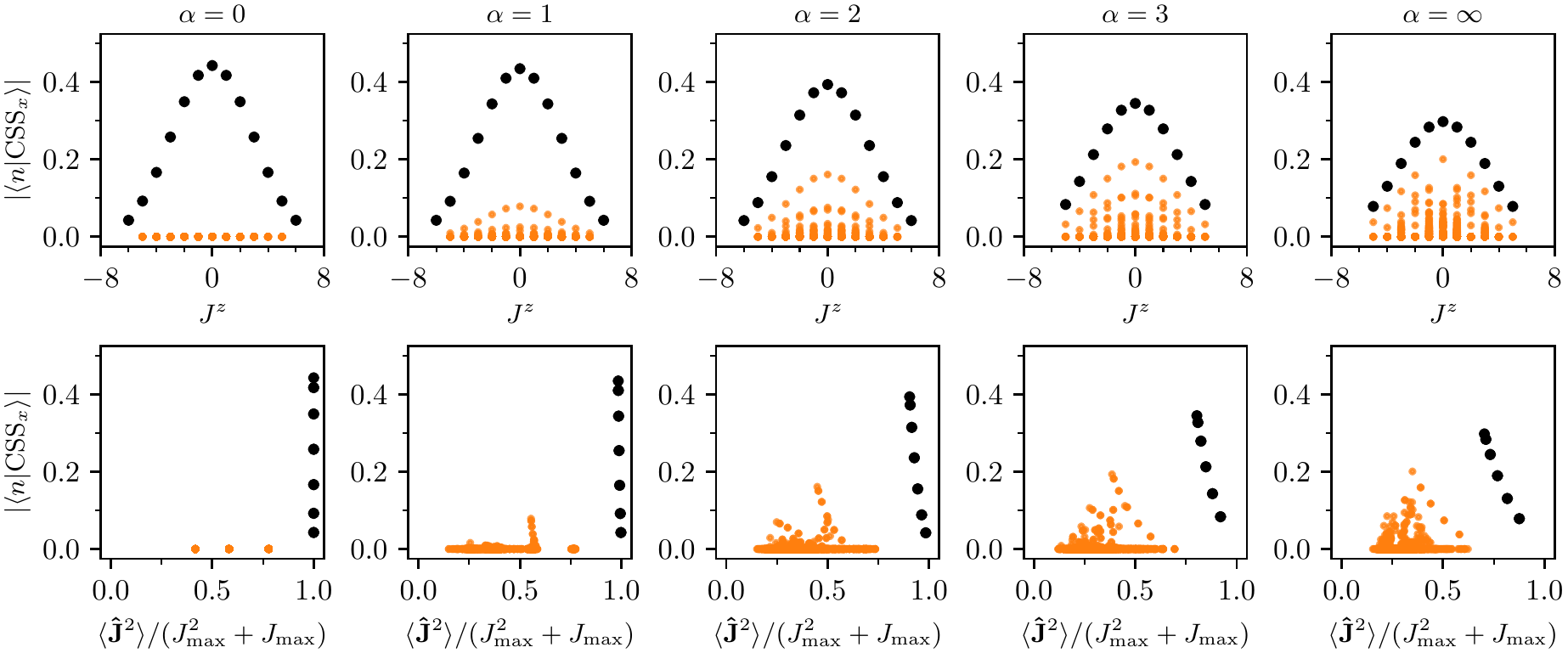}
\caption{
Overlap of the low-energy eigenstates of the $\alpha-XX$ model (with $d=1$ and $N=16$) with $|{\rm CSS}_x\rangle$, as a function of either $(J^z)^2$ (top panels) or the total angular momentum (lower panel).
Black points are states in the ToS, while orange (gray) dots are for the rest of the spectrum.
}
\label{sf.overlaps}
\end{figure*}

\section{Overlap of Coherent Spin State with the Tower of States}
\label{app:overlap}

Through exact diagonalization, we obtain the 3000 lowest-energy eigenstates $|n\rangle$ of the $\alpha-XX$ Hamiltonian $\hat{\cal{H}}$ (in $d = 1$ and with $N = 16$).
Fig. \ref{sf.overlaps} shows that for $\alpha=0$ the eigenstates forming the ToS are the only ones with a finite overlap $|\langle n | {\rm CSS}_x\rangle|$ with the CSS. As $\alpha$ increases, also other states in the spectrum acquire a finite overlap.

{
\begin{figure*}[hbt!]
\centering
\includegraphics[width=0.9\linewidth]{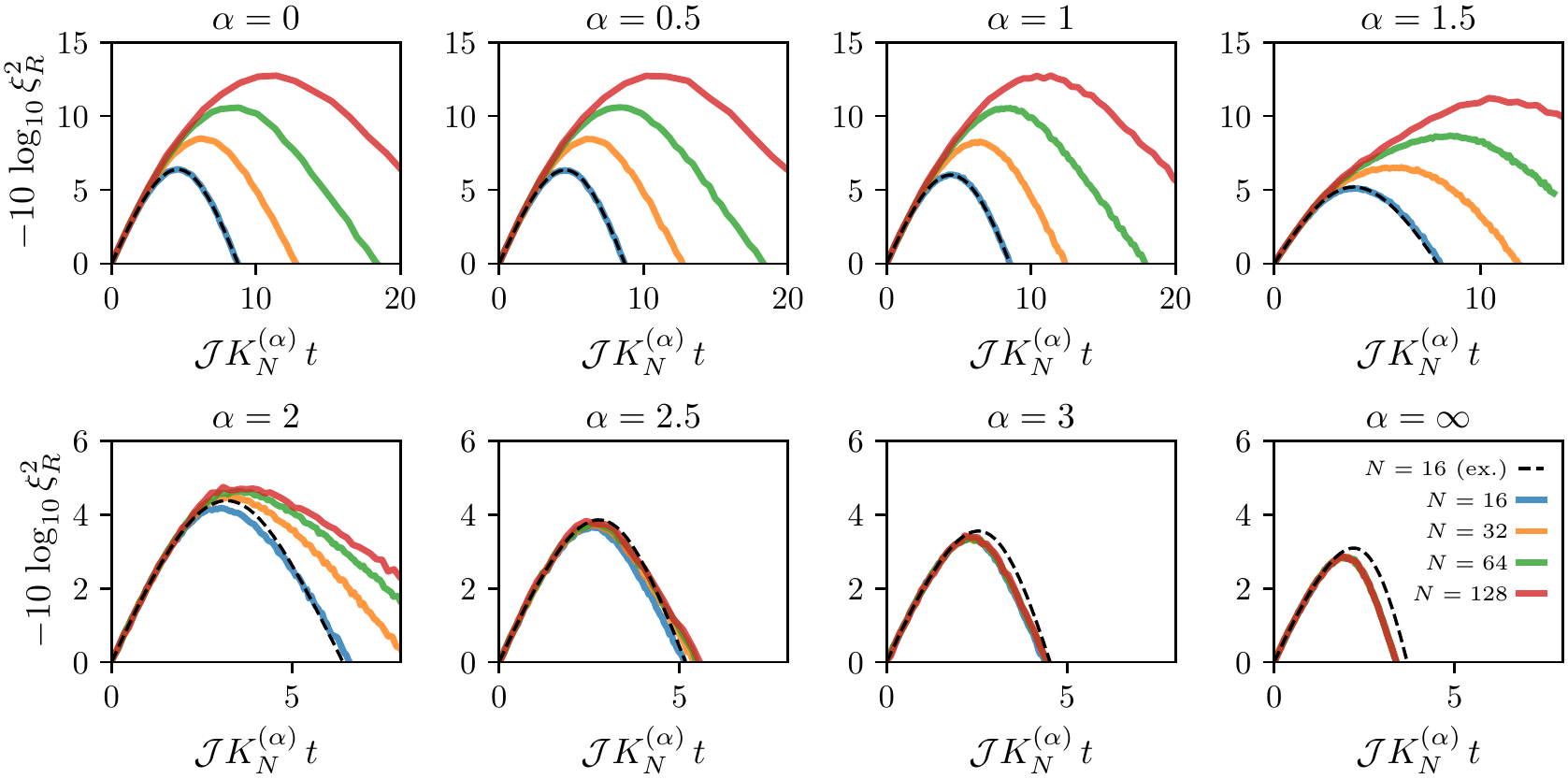}
\caption{
Time evolution of spin squeezing generated by the $\alpha-XX$ Hamiltonian, with $d=1$ and $\alpha\in\lbrace 0, 1/2, 1, 3/2, 2, 5/2, 3, \infty \rbrace$.
Results are obtained via tVMC with the pair-product Ansatz (solid lines) for system sizes from $N=16$ (bottom line) to $N=128$ (top line). For each $\alpha$ we also show the exact result for $N=16$ (dashed black line). Time is rescaled by the Kac prefactor. Statistical error bars on the tVMC curves are of the order of the line width or smaller.
}
\label{sf.squeezing_more_cases_kac}
\end{figure*}

\section{Squeezing dynamics for different interaction ranges}
\label{app:squeezing_data}

Here we show additional numerical results for the spin-squeezing dynamics, for several values of $\alpha$ -- see Fig.~\ref{sf.squeezing_more_cases_kac}. For $N=16$, we compare the squeezing dynamics obtained through tVMC with the exact one, and we observe that the accuracy of the pair-product Ansatz improves when $\alpha$ decreases (becoming exact at $\alpha = 0$).
As we consider the Kac-normalized $\alpha-XX$ model, which corresponds to multiplying time by $K_N^{(\alpha)}$, curves for different system sizes collapse onto each other at short times -- for any value of $\alpha$. The dependence on system size appears at larger time, where we observe the presence or absence of scaling of the optimal squeezing with $N$. {The scaling analysis of the maximum squeezing and of the corresponding optimal squeezing time for the data of Fig.~\ref{sf.squeezing_more_cases_kac} is presented in Fig.~\ref{f.squeezing_scaling}. }

\section{Diagonal part of the $\alpha-XX$ Hamiltonian on the $|J_{\rm max}, J^z\rangle$ reduced basis}
\label{app:diagonal}

Fig.~\ref{sf.reduced_Ham} shows the diagonal elements of the 1d $\alpha-XX$ Hamiltonian with $N=16$ on the reduced basis $|J_{\rm max},J^z\rangle$. The Hamiltonian, commuting  with $\hat{J}^z$, is diagonal on this basis, and the diagonal matrix elements show a clear quadratic dependence on $J^z$ of the kind $\langle J_{\rm max}, J^z | \hat{\cal H} | J_{\rm max}, J^z\rangle = {\rm const.} + (J^z)^2/(2I^{(\alpha)})$, from which we can extract systematically the moment of inertia $I^{(\alpha)}$ shown in the main text. 

\begin{figure}[hbt!]
\centering
\includegraphics[width=0.8\linewidth]{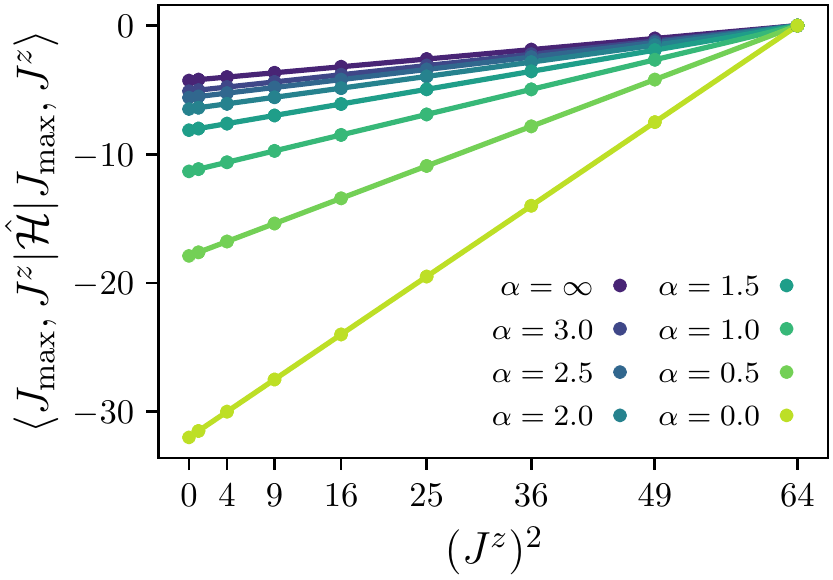}
\caption{Diagonal matrix elements of the $\alpha-XX$ Hamiltonian $\hat{\mathcal{H}}$ (for $d=1$ and $N=16$) on the $|J_{\rm max}, J^z \rangle$ states, as a function of $(J^z)^2$. For each $\alpha$, straight lines are linear fits. The exponent $\alpha$ ranges from $\alpha=\infty$ (top) to $\alpha=0$ (bottom).}
\label{sf.reduced_Ham}
\end{figure}

\section{Time evolution of the total spin}
\label{app:Jsquare}

For any $\alpha > 0$, the total spin $\hat{\bm J}^2$ has nontrivial dynamics, since it does not commute with the Hamiltonian $\hat{\cal H}$.
In the initial state $|{\rm CSS}_x\rangle$, $\langle \hat{\bm J}^2 \rangle$ is maximum and equal to $\langle \hat{\bm J}^2 \rangle(0) = J_\mathrm{max} (J_\mathrm{max} + 1)$. During the dynamics, for $\alpha > 0$, the $J=J_{\rm max}$ sector is mixed with other sectors, and $\langle \hat{\bm J}^2 \rangle (t)$ slowly departs from its initial value -- see Fig. \ref{sf.Jsq}. The decay is stronger the larger $\alpha$, due to the stronger $\hat{\cal H}_{q\neq 0}$ terms in the Hamiltonian.

\begin{figure*}[hbt!]
\centering
\includegraphics[width=0.95\linewidth]{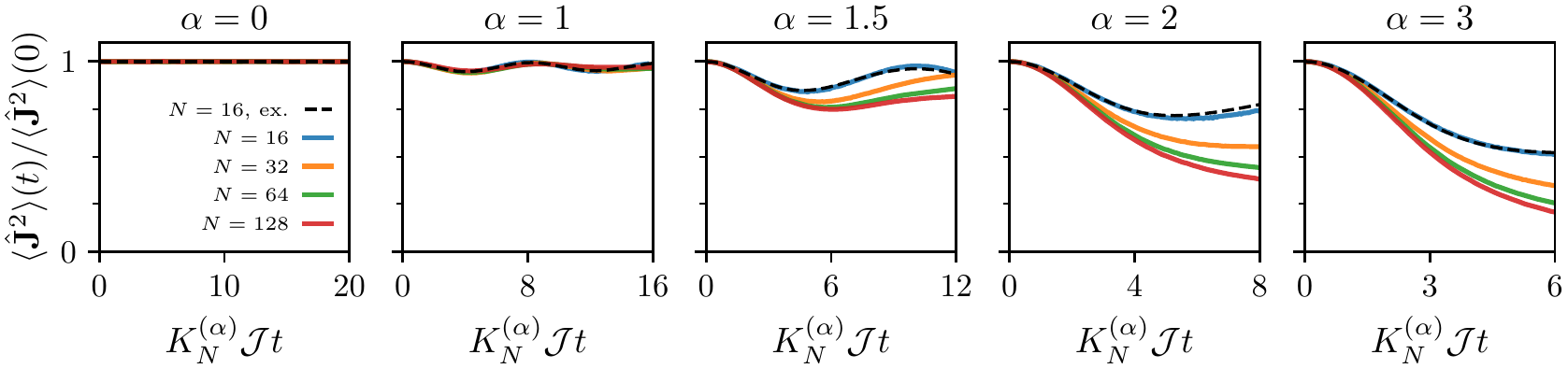}
\caption{Dynamics of the total spin $\langle \hat{\bm J}^2 \rangle$ in the one-dimensional $\alpha-XX$ model, for several values of $\alpha$. For each $\alpha$, tVMC results (solid lines) are shown for system sizes from $N=16$ (top line) to $N=128$ (bottom line), and compared with the exact curve for $N=16$ (dashed black line).}
\label{sf.Jsq}
\end{figure*}

\section{Spin squeezing dynamics of models not featuring the Anderson Tower of states}
\label{app:dimerized}

\begin{figure*}[hbt!]
\centering
\includegraphics[width=0.95\linewidth]{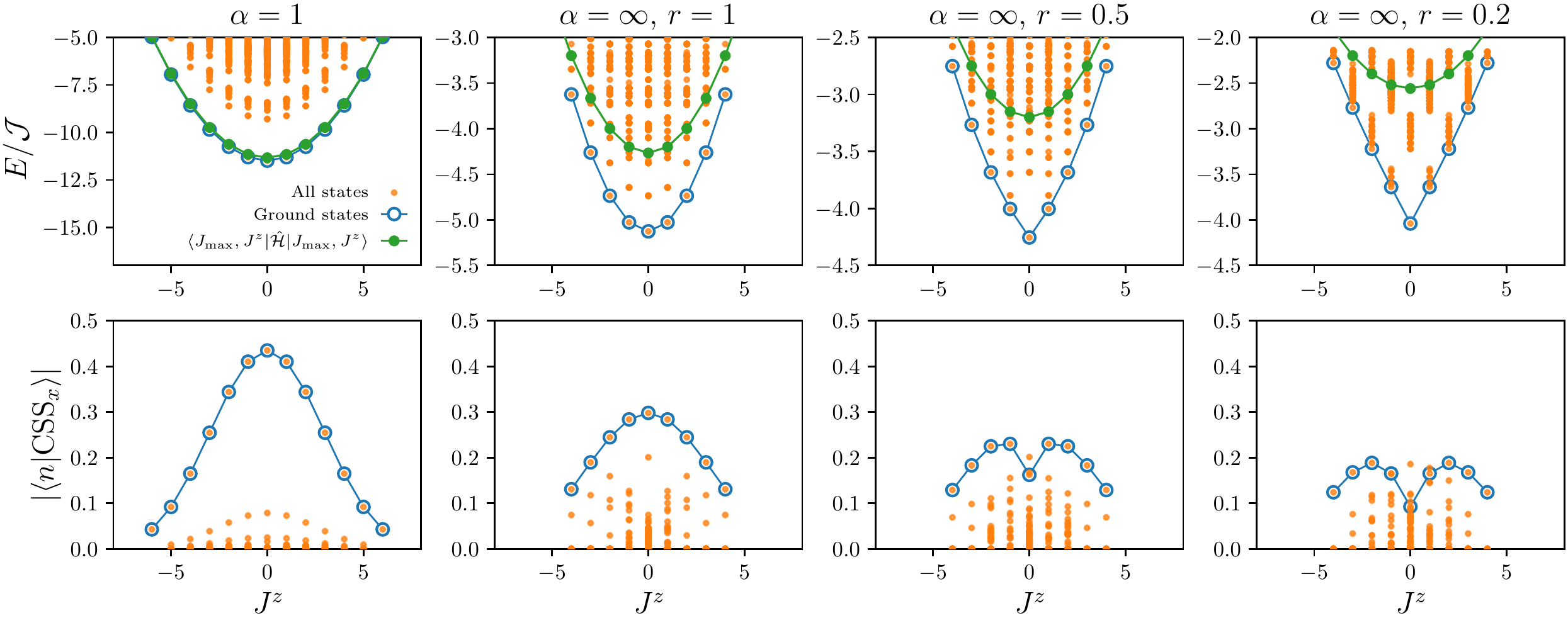}
\caption{\emph{Upper row.} $J^z$-resolved spectra of $\alpha-XX$ chains ($N=16$) with long-range interactions ($\alpha=1$), with short-range interactions ($\alpha=\infty$), and with dimerized short-range interactions ($r={\cal J}'/{\cal J} < 1$). The blue (empty) circles highlight the ground states in each $J^z$ sector, while the green (filled, dark gray) dots indicate the spectrum of the Dicke-projected Hamiltonian, $E_{\rm Dicke}(J^z) = \langle J_{\rm max}, J^z | \hat {\cal H} | J_{\rm max}, J^z\rangle$.  \emph{Lower row.} $J^z$-resolved overlap of the Hamiltonian eigenstates with the CSS. Symbols are the same as in the upper row. For $\alpha=\infty$, slight asymmetries are visible between the positive and negative $J^z$ values: they are an artifact of exact diagonalization mixing degenerate eigenstates.}
\label{sf.dim_spectra}
\end{figure*}

To further stress the role of the ToS in squeezing dynamics, we examine here the spectral features and squeezing dynamics of a class of models \emph{not} featuring an Anderson ToS. Such a class is represented by dimerized $\alpha$-$XX$ chains with nearest neighbor interactions ($\alpha = \infty$):
\begin{equation}
\begin{aligned}
{\hat{\cal H}} =&
- {\cal J} \sum_{n=1}^{N/2} (\hat{S}^x_{2n} \hat{S}^x_{2n+1} + \hat{S}^y_{2n} \hat{S}^y_{2n+1})\\
&
- {\cal J}' \sum_{n=0}^{N/2-1} (\hat{S}^x_{2n+1} \hat{S}^x_{2n+2} + \hat{S}^y_{2n+1} \hat{S}^y_{2n+2})
\end{aligned}
\end{equation} 
featuring different couplings, ${\cal J}$ and ${\cal J}'$, for even and odd bonds, respectively. In the following we shall take $r = {\cal J}'/{\cal J} < 1$ for definiteness. This model is equivalent to that of free fermions on a dimerized chain, or a Su-Schrieffer-Heeger model \cite{Su1979PRL} at half filling. 

\begin{figure}[ht!]
\centering
\includegraphics[width=0.8\linewidth]{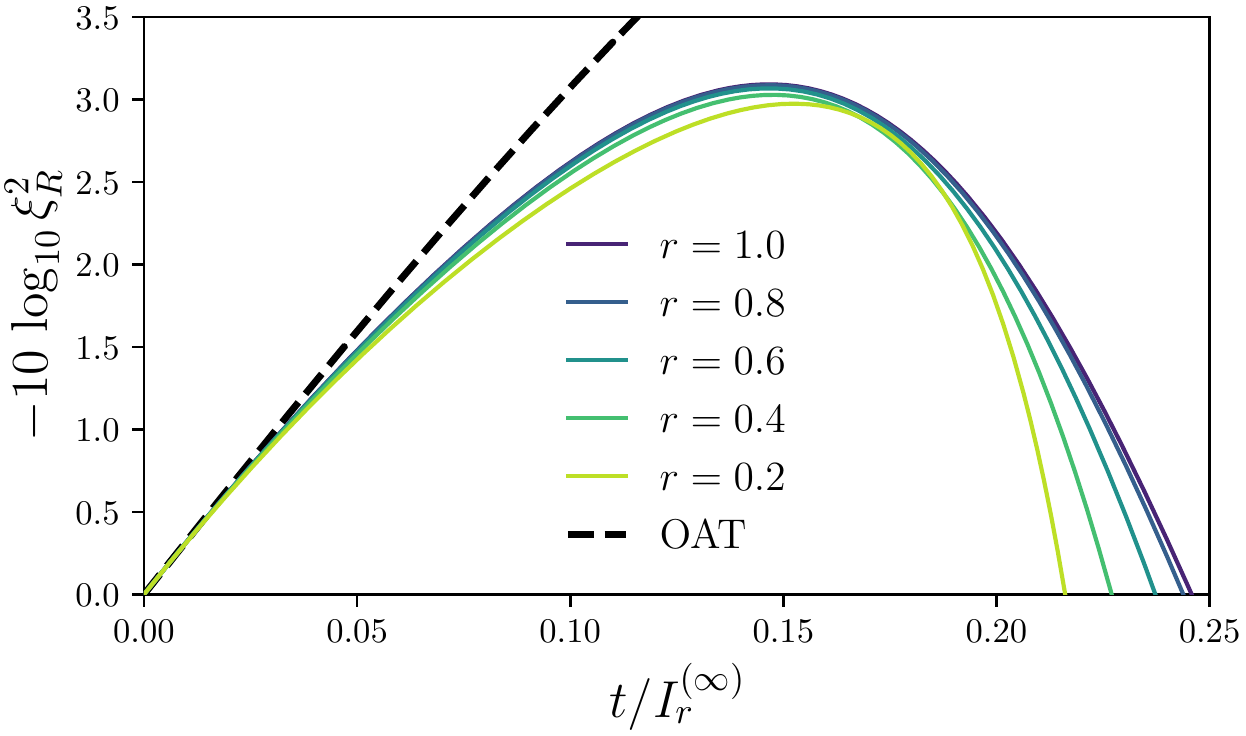}
\caption{Spin squeezing dynamics of an $N=16$ $XX$ chain with dimerization. The time axis is rescaled with the moment of inertia $I^{(\infty)}_r$ of the Dicke-projected spectrum, so that the  squeezing dynamics becomes identical to that of the OAT model at very short times. The $r$ parameter ranges from $r=0.2$ (lightest line) to $r=1$ (darkest line).}
\label{sf.dim_squeezing}
\end{figure}

Fig.~\ref{sf.dim_spectra} (first row) shows the $J^z$-resolved low-energy spectrum for the dimerized Hamiltonian with two values of ${\cal J}'$ (${\cal J}/2$ and ${\cal J}/5$), contrasted with the $\alpha=1$ and $\alpha=\infty$ case without dimerization. The same spectrum is compared with the dispersion relation of the Dicke-projected Hamiltonian, namely the energy spectrum 
$E_{\rm Dicke}(J^z) = \langle J_{\rm max}, J^z | \hat {\cal H} | J_{\rm max}, J^z\rangle$, whose quadratic dispersion $E_{\rm Dicke}(J^z)  \approx {\rm const.} +  (J^z)^2/(2I^{(\alpha)})$ dictates the short-time squeezing dynamics, as discussed in the main text.  In the second row of the same figure we show the $J^z$-resolved overlap of the Hamiltonian eigenstates with the initial state ($|{\rm CSS}_x\rangle$) of the squeezing evolution. In both panels the open circles single out the ground states in each $J^z$ sector. 

For long-range interactions ($\alpha = 1$ in Fig.~\ref{sf.dim_spectra}) the states composing the Anderson ToS are extremely close to Dicke states, so that the dispersion relation of the ToS is essentially the same as that of the Dicke-projected Hamiltonian. At the same time, the states of the ToS have  by far the largest overlap with $|{\rm CSS}_x\rangle$, and therefore they control the dynamics starting from $|{\rm CSS}_x\rangle$. On the other hand, even in the absence of dimerization, for short-range interactions ($\alpha = \infty$) the ToS features an overlap with the $|{\rm CSS}_x\rangle$ which is compatible with that of states outside of the ToS; consistently with this picture, the spectrum of the Dicke-projected Hamiltonian falls into the spectrum of the excited states above the ToS. Nonetheless, the curvature of the Dicke-projected spectrum is still close to that of the ToS, so that the two moments of inertia $I^{(\alpha)}$ and $I^{(\alpha)}_{\rm ToS}$ (as defined in the main text) for $\alpha = \infty$ remain close to each other (see Fig. 3(c) of the main text). 

 When turning on the dimerization ($r<1$), the spectrum of the Dicke-projected Hamiltonian remains fully quadratic in $J^z$, namely one can still define a moment of inertia $I^{(\infty)}_{r}$ such that $E_{\rm Dicke}(J^z)  \approx {\rm const.} + (J^z)^2/(2I^{(\infty)}_{r})$. As a consequence the dynamics of the model at very short times, dominated by the Dicke-projected Hamiltonian, still features squeezing similar to that of an OAT model with an effective moment of inertia $I^{(\infty)}_{r}$. This is indeed observed in Fig.~\ref{sf.dim_squeezing} showing that all dimerized Hamiltonians lead to short-time squeezing, in accordance with the theorem of Ref.~\cite{Roscilde2021PRA}. 
 
 But the similarity with the situation of systems without dimerization is only apparent. Indeed the spectrum of the ground states in each $J^z$ sector is no longer parabolic, but it acquires a linear term, namely $E_0(J^z)  \approx {\rm const.} + a |J^z| + b (J^z)^2$ -- as it is clearly visible in Fig.~\ref{sf.dim_spectra}. This is related to the vanishing of the ground-state susceptibility at zero field as a consequence of dimerization. Therefore, the notion of ToS for the low-lying states, namely of a rotor-like spectrum and an associated effective moment of inertia, is no longer meaningful. Concomitantly the low-lying states no longer have a special overlap with $|{\rm CSS}_x\rangle$. Hence the relationship between the low-energy spectral structure of the Hamiltonian and its squeezing dynamics is lost. }

\bibliography{refs.bib}

\end{document}